\documentclass{aa}  

\usepackage{graphicx}
\usepackage{subfigure}
\usepackage{txfonts}
\usepackage[dvipsnames]{xcolor}
\usepackage[switch]{lineno}
\usepackage{hyperref}
\definecolor{coolblack}{rgb}{0.0, 0.18, 0.39}
\definecolor{darkblue}{rgb}{0.0, 0.0, 0.55}
\definecolor{mediumred-violet}{rgb}{0.78, 0.08, 0.52}
\hypersetup{
    colorlinks=true,       % false: boxed links; true: colored links
    linkcolor=darkblue,
    filecolor=magenta,      
    urlcolor=darkblue,
    citecolor=darkblue 
}
\usepackage[dvipsnames]{xcolor}
\usepackage{tablefootnote}
\usepackage[separate-uncertainty=true]{siunitx}
\newlength\figureheight
\newlength\figurewidth
\DeclareSIUnit\year{yr}
\newcommand{\N}{NRAO\,150}

\newcommand{\new}[1]{\textcolor{black}{#1}}

\begin{document}

   % \title{Probing the $\gamma$-ray emission region and the connection to jet ejections in NRAO~150}
   % \title{Investigate the $\gamma$-ray radio jet connection in NRAO~150 with VLBI}
   % \title{Linking $\gamma$-ray emission to VLBI radio ejections in NRAO~150}
   \title{Probing the $\gamma$-ray emission region and the connection to jet ejections in NRAO~150 with VLBI}
    % Pinpointing the gamma-ray emission region in NRAO150

   \author{
   L.~C. Debbrecht \inst{\ref{mpifr}}\thanks{\email{ldebbrecht@mpifr-bonn.mpg.de} \newline Member of the International Max Planck Research School (IMPRS) for Astronomy and Astrophysics at the Universities of Bonn and Cologne} , %et al. 
   G.~F. Paraschos\inst{\ref{Finca},\ref{Mh},\ref{mpifr}}, 
   E. Ros\inst{\ref{mpifr}},
   I. Agudo\inst{\ref{IAA}},
   T.~P. Krichbaum\inst{\ref{mpifr}},
   H. Müller\inst{\ref{Socc}, \ref{mpifr}},
   S.~G. Jorstad\inst{\ref{BU}, \ref{RU}},\\
   A.~P. Marscher\inst{\ref{BU}},
   M.~A. Gurwell\inst{\ref{CfA}}, %mgurwell@cfa.harvard.edu
   J.~A. Zensus\inst{\ref{mpifr}}
          }
    \authorrunning{L.~C. Debbrecht et al.}
   \institute{
         Max-Planck-Institut für Radioastronomie, Auf dem Hügel 69, D-53121 Bonn, Germany\label{mpifr}
          \and
          Finnish Centre for Astronomy with ESO, University of Turku, FI-20014 Turku, Finland\label{Finca}\and
          Aalto University Metsähovi Radio Observatory, Metsähovintie 114, FI-02540 Kylmälä, Finland\label{Mh}\and
          Instituto de Astrofísica de Andalucía, IAA-CSIC, Glorieta de la Astronomía s/n, E-18008 Granada, Spain \label{IAA}\and
          % National Radio Astronomy Observatory, Charlottesville, Virginia, USA \label{NRAO}\and
          National Radio Astronomy Observatory, PO Box O, Socorro, NM 87801, USA\label{Socc}\and
          Institute for Astrophysical Research, Boston University, 725 Commonwealth Avenue, Boston, MA 02215, USA \label{BU}\and
          Saint Petersburg State University, 7/9 Universitetskaya nab., St. Petersburg, 199034 Russia \label{RU}\and
          Center for Astrophysics | Harvard \& Smithsonian, 60 Garden Street, Cambridge, Massachusetts, 02138 USA\label{CfA}%\and
             }
   \date{Received XX; accepted YY}
% \abstract{}{}{}{}{} 
% 5 {} token are mandatory
  \abstract{
   Relativistic jets launched by active galactic nuclei are fundamental for understanding the physics of accreting supermassive black holes and their immediate environments, yet the origin of these jets remains an open question.
   NRAO 150 is a blazar with a complex relativistic jet morphology that evolves on short timescales due to strong projection effects, enabling detailed kinematic analysis.
   In this study, we utilise data by the Very Long Baseline Array and the European VLBI Network from 2010 until 2019 at 43\,GHz, to understand the formation and launching processes of the jet in NRAO 150. 
   We study the $\gamma$-ray and radio light-curves, together with total intensity and linear polarisation information to probe the connection between flaring events, $\gamma$-ray emission, and the ejection of new jet features. 
   Furthermore, we investigate the magnetic field configuration in the innermost jet region, as captured in polarised light, to gain insights about its configuration before, during, and after a $\gamma$-ray flare. 
   Our results indicate a close temporal link between the $\gamma$-ray flaring activity and the ejection of new VLBI jet components, suggesting that the high-energy emission is produced downstream of the VLBI core. 
   The combined kinematic and polarimetric evidence further points to a toroidal magnetic field in the inner jet, highlighting the key role of magnetic fields in governing both jet dynamics and high-energy emission in NRAO 150.
   }
   \maketitle
%
%-------------------------------------------------------------------
\section{Introduction}
Blazars are the most luminous type of Active Galactic Nuclei (AGN), with their energetic jets pointing at a small angle towards our line of sight (LOS). 
These objects are perfectly suited to study their kinematic behaviour, as their morphology changes on short timescales due to projection effects. 
Understanding the processes and mechanisms driving their powerful jets is crucial to investigate how their jets are launched and their connection to the central supermassive black hole (SMBH). 
One such blazar is \N\, at a redshift of 1.52 \citep{Acosta}, whose jet is viewed at a small angle towards our line of sight; estimates of the viewing angle range from a few degrees ($\theta\approx \SI{8}{\degree}$) down to values approaching $\theta\sim \SI{0}{\degree}$ \citep{Agudo2007}. 
This blazar is of particular interest, as its viewing angle enhances relativistic beaming of individual emission features, facilitating detailed studies of their kinematic behaviour. 
Very long baseline interferometry (VLBI) at millimetre wavelengths enables the jet emission to be resolved into discrete components, commonly interpreted as moving shocks or regions of enhanced synchrotron emission~\citep[see][for a review]{Boccardi2017}.
Multi‑epoch monitoring of such components provides a powerful tool to constrain jet kinematics and to investigate their temporal relation to high‑energy activity.

% ---------------------
%     Morphology: Kinematics and polarisation/ magnetic field config 
% ---------------------
Magnetic fields are expected to play a fundamental role in both jet dynamics and radiative processes.
Enhanced linear polarisation in AGN jets is commonly interpreted as a signature of increased ordered magnetic fields, and may be associated with episodes of enhanced particle acceleration and flaring activity \citep[][and references therein]{Laing1980, GP2024EHT, GP2026}.
Polarimetric analysis by \cite{Molina2014} and \cite{Livingston2025}, using the very long baseline array (VLBA) and global mm-VLBI array (GMVA), point to a toroidal and helical magnetic field threading the jet of \N. 
% ---------------------
%       Gamma 
% ---------------------
In blazars, $\gamma$-ray flares are often temporally associated with the ejection of new VLBI components, suggesting a close link between high-energy emission and the evolution of the relativistic jet~\citep{Jorstad2001}. 
However, the exact location of the $\gamma$-ray emission site remains a subject of investigation. 
Previous studies suggest that the site of high-energy emission is located jet downstream from the VLBI core in the extended radio jet~\citep{Agudo2011, Agudo2011b, GP2025a, GP2026}. 
\N\, has been in an active $\gamma$‑ray state since July 2013, making it an excellent target for investigating the connection between $\gamma$‑ray flaring activity, jet component ejections, and magnetic field structure.
\cite{Zhou2018} conducted temporal analysis of the $\gamma$-ray light-curve and found, that \N\, shows a variability of several hours, which is not common for high-redshift blazars. 
They analysed the emission region of the $\gamma$-rays, which size is comparable to the Schwarzschild radius~\citep[i.e. $4.79\times 10^{-4}$pc; ][]{Acosta}. 

Until today, the exact physical processes explaining the aforementioned kinematic behaviour and high-energy emission are not yet fully understood. 
In this work we present a multi-epoch analysis of \N\, using millimetre VLBI observations, combining high-resolution total intensity and linear polarisation images, together with radio and $\gamma$-ray light-curve information. 
We aim to investigate the spatio-temporal relationship between $\gamma$-ray flares and the ejection of new jet components, and to constrain the location of the high-energy emission site within the relativistic jet of \N. 
Studying the linear polarisation information, we further probe the magnetic field configuration and its possible connection to the high-energy emission. 

This paper is structured as follows: In Sect.~\ref{Sec:ObsDataRed} we provide information about the observations and data reduction. 
Sect.~\ref{Sec:MethodsResults} outlines the methods and presents the results. 
In Sect.~\ref{Sec:Discussion} we discuss our results and in Sect.~\ref{Sec:Conclusion} we draw our conclusions of our analysis. 
Throughout this paper we assume a $\Lambda$ cold dark matter cosmology with $H_0 = 67.8\,\mathrm{kms}^{-1}\,\mathrm{Mpc}^{-1}$, $\Omega_\Lambda = 0.692$, and $\Omega_\mathrm{m} = 0.308$~\citep{PlanckCollab}, so that 1\,mas corresponds to 9\,pc, and 1\,mas/yr corresponds to 29.34\,$c$ at the redshift of \N. 

%--------------------------------------------------------------------
%--------------------------------------------------------------------
%--------------------------------------------------------------------
%--------------------------------------------------------------------
%--------------------------------------------------------------------
%--------------------------------------------------------------------
\section{Observations and data reduction \label{Sec:ObsDataRed}}
\N\, was observed by the VLBA-BU Blazar Monitoring Program (BEAM-ME and VLBA-BU-BLAZAR\footnote{\url{http://www.bu.edu/blazars/BEAM-ME.html}}) in 2010 until 2019 at 43\,GHz, and data by the European VLBI Network (EVN) in June 2014 and 2015 at the same frequency. 
The data were recorded in eight baseband channels with two-bit quantisation, with recording rate of 2\,Gbps. 
Finally, the data by the VLBA were calibrated using NRAO's Astronomical Image Processing System \citep[AIPS;][]{AIPS2003} with the procedure applied as described in~\cite{Park2021} and~\citeyearpar{Park2024}. 
The data of 2014 and 2015 were correlated and calibrated at the correlator at Joint Institute for VLBI European Research Infrastructure Consortium (JIVE) in Dwingeloo, Netherlands. 
The data of 2014 and 2015 were first averaged over all intermediate frequencies (IFs) and then time-averaged in 30-second bins after which we employed a hybrid imaging technique, iteratively combining the \texttt{CLEAN} deconvolution algorithm \citep{CLEANHogbom} in \texttt{DIFMAP}~\citep{Difmap} along with self-calibration in phase and amplitude. 
We applied a constant systematic non-closing error of 3\% to the data of 2014 and 2015, to address uncertainties in the absolute amplitudes of the visibilities and to account for low signal-to-noise measurements~\citep[see][for a detailed description]{EHT2019}. 
Table~\ref{tab:obs} summarises all information about the observations and the \textsc{clean}-maps of \N. 

In addition, we used polarisation information for all epochs. 
The polarisation calibration for the VLBA data was done as described in~\cite{Jorstad2017PolCalBU}, while the calibration of the EVN data was done following ~\cite{homanwardle1999} and ~\cite{Homan2001}. 
Discrepancies between the RR- and LL-correlations were addressed using a gain-transfer technique to correct for an instrumental RL gain offset, which is degenerate with intrinsic circular polarisation (CP). 
Assuming that the ensemble-averaged CP across multiple sources vanishes, a source-independent RL offset was derived by jointly calibrating all sources in the observing block.
The procedure was applied independently to each IF, with problematic datasets flagged as necessary. 
The calibration of the absolute position of the electric vector position angle (EVPA) was performed using single-dish observations from the Effelsberg 100\,m telescope, incorporating the rotation measure (RM) reported by \cite{Livingston2025}, and applying the EVPA relation described in~\cite{Hovatta2012}. 
Further, we incorporated $\gamma$-ray flux light-curves, that are publicly available from the Fermi Large Area Telescope Collaboration repository\footnote{\url{https://fermi.gsfc.nasa.gov/ssc/data/access/lat/LightCurveRepository/}}~\citep[Fermi-LAT; see][for a detailed description]{Atwood2009,Abdollahi2023FERMI}, and adopted a monthly binning to ensure sufficient photon statistics. 
We utilised radio flux measurements at 1.3\,mm of \N\, by the Submillimeter Array ~\citep[SMA\footnote{\url{http://sma1.sma.hawaii.edu/callist/callist.html}};][]{SMAHo2004,Gurwell2007}, which are publicly available data.

\begin{table*}
\caption{Summary of VLBI observations of \N\, at 43\,GHz. }
\centering
\begin{tabular}{cccccccc}
\hline 
Epoch &  Interferometer & Beam size & Beam position angle & $I_\mathrm{peak}$ & $\sigma_\textrm{I}$ & $I_\mathrm{P}$ & $m_\mathrm{P}$\\
\text{[yyyy-mm-dd]} & & [mas] & [$^\circ$] & [Jy/beam] & [mJy/beam] & [Jy] & [\%]\\ 
(1)\label{tab:data} & (2)\label{tab:freq} & (3)\label{tab:Intf} & (4)\label{tab:beam} & (5)\label{tab:PA} & (6)\label{tab:Ipeak} & (7) & (8)\\
\hline\hline
2010-04-15 & VLBA & $0.35 \times 0.17$ & $-47.3$ & 3.5 & 1.2 & $0.36 \pm 0.05$ & $3.91 \pm 0.59$\\
2010-11-12 & VLBA & $0.34 \times 0.19$ & $-1.5$ & 2.8 & 1.2 & $0.20 \pm 0.03$ & $2.42 \pm 0.36$\\
2011-09-24 & VLBA & $0.33 \times 0.22$ & 32.0 & 2.7 & 1.7 & $0.26 \pm 0.04$ & $4.59 \pm 0.69$\\
\hline
2014-06-07 & EVN & $0.14 \times 0.09$ & $-0.3$ & 1.3 & 1.2  & $0.33 \pm 0.05$ & $9.56 \pm 1.43$\\
2015-06-01 & EVN & $0.14 \times 0.10$ & $-1.1$ & 1.2 & 1.2 & $0.28 \pm 0.04$ & $7.51 \pm 1.13$\\
\hline
2017-05-13 & VLBA & $0.26 \times 0.17$ & $-4.3$ & 1.3 & 4.1 & $0.23 \pm 0.03$ & $4.97 \pm 0.74$\\
2018-02-17 & VLBA & $0.35 \times 0.27$ & $-56.9$ & 1.1 & 3.0 & $0.30 \pm 0.04$ & $8.04 \pm 1.21$\\
2019-03-31 & VLBA & $0.28 \times 0.17$ & 10.0 & 1.2 & 0.7 & $0.14 \pm 0.02$ & $3.44 \pm 0.52$\\
 \hline 
\end{tabular}
\tablefoot{(1) Date of observation in year-month-day format. (2) Interferometer. (3) The nominal restoring beam sizes of the clean image in mas  (uniform weighting) and (4) the position angle of the major axis of the beam. (5) Total intensity peak of Stokes I in units of Jy/beam (using a circular beam of 0.25\,mas). (6) Total intensity RMS level in mJy/beam. (7) Integrated linear polarisation\footnote{We define the integrated linear polarisation as $I_\mathrm{P}=\Sigma_i |P_i|$, where summation of linear polarisation $P$ is over pixel $i$.} in Jy. \new{(8) Integrated fractional polarisation\footnote{\new{We define the integrated fractional  polarisation as $m_\mathrm{P}=\Sigma_i |P_i|$, where summation of fractional polarisation $P$ is over pixel $i$.}} in \%. } } 
\label{tab:obs}
\end{table*}

%--------------------------------------------------------------------
%--------------------------------------------------------------------
%--------------------------------------------------------------------
%--------------------------------------------------------------------
%--------------------------------------------------------------------
%--------------------------------------------------------------------
\section{Methods and results \label{Sec:MethodsResults}}
 
\begin{figure*}
    \centering
    \includegraphics[width=\textwidth]{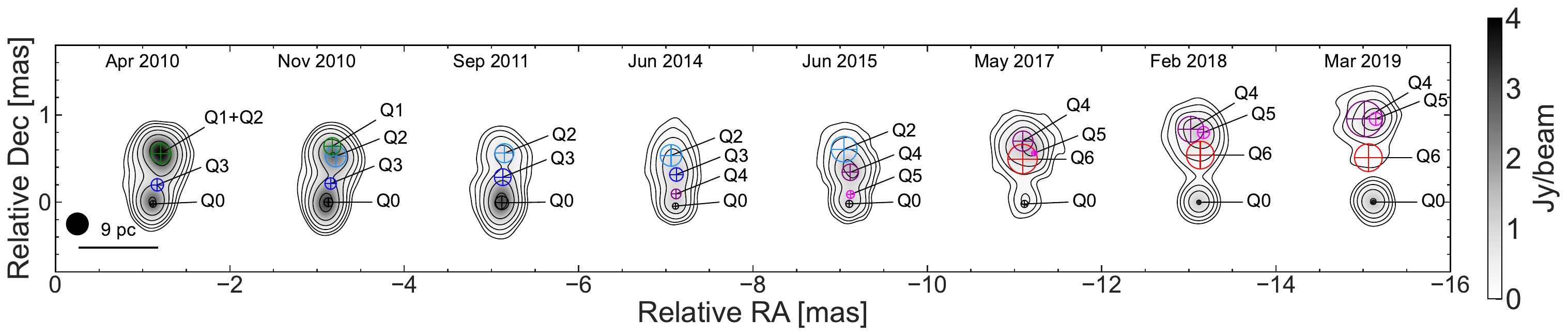}
    \caption{Total intensity images of \N\, of all epochs at 43\,GHz. 
    The total intensity flux density is represented by the grey colour scale and the contours, using the contour levels at 2, 4, 8, 16, 32, and 64\% of the peak flux. 
    The cut-off is at $7~\sigma_\textrm{I}$ for all epochs (with $\sigma_\textrm{I} = 2.3~\textrm{mJy/beam}$). 
    The black circle in the bottom left corner denotes the convolving, circular beam size of 0.25\,mas for all epochs and the black dash in the bottom left corner denotes the projected distance of $9\,\mathrm{pc}$ corresponding to $20000~R_\mathrm{S}$. 
    The fitted model-fits are colour coded. 
    }
    \label{fig:Modelfitmaps}
\end{figure*}
\begin{figure*}
    \centering
    \includegraphics[width=\textwidth]{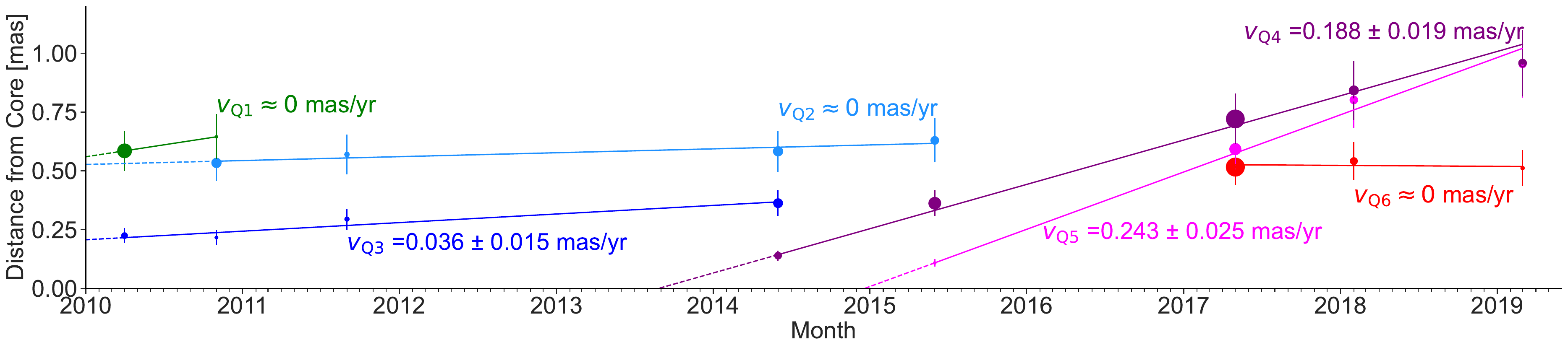}
    \caption{Plot of the cross-identified and colour-coded emission features in projected distance from the VLBI core in mas versus epochs in months in \N.
    The size of each circle is normalised to the flux density of the central core model-fit component of that epoch. 
    The solid lines correspond to the linear fit and the dashed lines correspond to the extrapolated intersection with the x-axis. 
    }
    \label{fig:Dist}
\end{figure*}

\subsection{Kinematic analysis}
The Stokes~I images of \N\, at 43\,GHz are presented in Fig.~\ref{fig:Modelfitmaps}, using a common circular convolving beam size of $0.25~\mathrm{mas}$. 
In all epochs the jet is directed to the north extending up to $\sim1.2~\mathrm{mas}$. 
The colour coded circles indicate \textsc{model-fit} components to identify moving emission features, and to monitor their kinematics across all epochs. 
\textsc{model-fit} components are fitted circular Gaussians to the visibilities, of which the sizes are normalised to the core component (black). 
We denote the core component as Q0 and all other components in the jet downstream as Q1-Q6. 
In previous analyses, such as~\cite{Agudo2007} and~\cite{Molina2014}, an additional eastern component was observed in earlier epochs (2006-2009), which was not detected in the epochs presented here. 
We tested the fit of an additional component within the \textsc{model-fit} procedure but this did not result in a reasonably reduced $\chi^2$. 
Figure~\ref{fig:Dist} displays the projected distance of the \textsc{model-fit} components to the core component in milliarcseconds (mas) versus time, in which the colours of the data points correspond to the colour-coded \textsc{model-fit} components in Fig.~\ref{fig:Modelfitmaps}. 
We adopted an uncertainty of 15\% on the positions of the \textsc{model‑fit} components, using a conservative estimate that accounts for both the angular resolution and uncertainties from the \textsc{model‑fit} procedure~\citep{Punsly2021}. 

Further, we determined the projected velocities of each \textsc{model-fit} component by applying a linear fit $d(t)= v_0t+e$, in which $d$ is the projected distance from the core to the \textsc{model-fit} component in mas, $t$ is the epoch in yr, $v_0$ is the projected velocity in mas/yr, and $e$ is the ejection epoch. 
The linear fit is represented by the solid line in Fig.~\ref{fig:Dist}, and the dashed line denotes the extrapolated the fit to estimate the time of ejection, while we assume no acceleration of the emission features. 
The information about the fit parameters and all \textsc{model-fit} components are listed in Tab.~\ref{tab:kinematics} and Tab.~\ref{tab:components}, respectively. 
The identification and cross-referencing of components across multiple epochs was determined by the smallest possible $\chi^2$ of the linear fit 
\new{and by demanding 
% In addition, the cross-identification was verified by the 
continuity in the observed component properties across epochs, in particular the flux density and full width half maximum (FWHM) of the \textsc{model-fit} components (see Tab.~\ref{tab:components}). 
These quantities evolve smoothly for the identified features and support the adopted component associations.
We note, that the FWHM of feature Q1+Q2 is comparatively large (see Fig.~\ref{fig:Modelfitmaps}), which may indicate that two components are overlapping and are therefore too close to each other in order to individually resolve them. 
% can be explained by the fact that these components may be overlapping and are therefore too too close to each other in order to resolve them. 
}
% We note, that the FWHM of feature Q1+Q2 is comparatively large (see Fig.~\ref{fig:Modelfitmaps}), which can be explained by the fact that these components may be overlapping and are therefore too too close to each other in order to resolve them. }
% We note, that the FWHM of component Q1+Q2 is comparatively large (see Fig.~\ref{fig:Modelfitmaps}), which can be explained by the fact that this component may be also containing component Q2, and therefore appearing larger. 
Further, component Q6 appears to be a stationary feature, which may be due to the absence of additional epochs after 2019, meaning that we can not find subsequent component to connect to. 
Component Q3, however, moves with a projected velocity of $\sim0.036\pm0.015~\mathrm{mas/yr}$. 
In 2014 and 2015, two new components were ejected (Q4 and Q5), with velocities at least $\sim5$ times higher than in the previous years, that are $\sim0.188\pm0.019~\mathrm{mas/yr}$ and $\sim0.243\pm0.025~\mathrm{mas/yr}$. 
These velocities correspond to superluminal apparent speeds of $1.06 \pm 0.44\,c$ (Q3), $5.52 \pm 0.56\,c$ (Q4), and $7.13 \pm 0.73\,c$ (Q5). 
% The analysis of the ejection epoch of each component 
We find that component Q3 was ejected in $2004.30\pm903\,\mathrm{days}$ (corresponding to April 2004), Q4 was ejected in $2013.67 \pm217\,\mathrm{days}$ (August 2013), and component Q5 in $2014.92\pm269\,\mathrm{days}$ (December 2014). 
The interpretation and implications of these observations are discussed in Sect.~\ref{subsec:Kin}. 

\begin{table*}[h]
\caption{Linear-fit kinematic parameters of jet components.}
\centering
\begin{tabular}{ccccc}
\hline
Component & Projected speed & Ejection epoch & $\sigma_\mathrm{ej}$ & ${\beta_\mathrm{app}}^\dagger$ \\ 
 & [$\mathrm{mas/yr}$] &  & [days] & [$c$]\\
(1) & (2) & (3) & (4) & (5)\\
\hline\hline
Q1 & $\sim0 $ & - & - & -\\
Q2 & $\sim0 $ & - & - & -\\
Q3 & $0.036 \pm 0.015$ & 2004.30 (Apr 2004) & $903$ & $1.06 \pm 0.44$\\
Q4 & $0.188 \pm 0.019$ & 2013.67 (Aug 2013) & $217$ & $5.52 \pm 0.56$\\
Q5 & $0.243 \pm 0.025$ & 2014.92 (Dec 2014) & $269$ & $7.13 \pm 0.73$\\
Q6 & $\sim0 $ & - & - & - \\
\hline
\end{tabular}
\tablefoot{
Column (1): Component identifier.
Column (2): Projected separation speed.
Column (3): Ejection epoch estimated from the zero-separation intercept.
Column (4): Uncertainty of the ejection epoch expressed in days.
Column (5): Apparent proper motion in units of the speed of light, $^\dagger \beta_\mathrm{app}= \frac{\beta \sin\theta}{1-\beta\cos\theta}$, in which $\beta$ is the speed in units of the speed of light $c$ and $\theta$ is the angle between the direction of the flow and the line of sight. 
}
\label{tab:kinematics}
\end{table*}

\begin{table*}
\caption{Information about \textsc{model-fit} components. }
\centering
\begin{tabular}{ccccccc}
\hline
Component & Epoch & Flux density & FWHM & Distance from core & PA \\
 &  & [Jy] & [mas] & [mas] & [$^\circ$] \\
(1) & (2) & (3) & (4) & (5) & (6) \\
\hline\hline
Q0 & Apr 2010 & $2.36 \pm 0.35$ & $0.08 \pm 0.01$ & $0.00 \pm 0.00$ & $0.00 \pm 0.00$ \\
Q3 & Apr 2010 & $1.05 \pm 0.16$ & $0.14 \pm 0.02$ & $0.22 \pm 0.03$ & $-12.78 \pm 2.62$ \\
Q1+Q2 & Apr 2010 & $5.95 \pm 0.89$ & $0.23 \pm 0.04$ & $0.59 \pm 0.09$ & $-8.76 \pm 1.83$ \\
\hline
Q0 & Nov 2010 & $3.61 \pm 0.54$ & $0.10 \pm 0.02$ & $0.00 \pm 0.00$ & $0.00 \pm 0.00$ \\
Q3 & Nov 2010 & $0.41 \pm 0.06$ & $0.13 \pm 0.02$ & $0.22 \pm 0.03$ & $-6.68 \pm 1.40$ \\
Q2 & Nov 2010 & $4.19 \pm 0.63$ & $0.24 \pm 0.04$ & $0.53 \pm 0.08$ & $-9.06 \pm 1.89$ \\
Q1 & Nov 2010 & $0.21 \pm 0.03$ & $0.20 \pm 0.03$ & $0.64 \pm 0.10$ & $-3.97 \pm 0.84$ \\
\hline
Q0 & Sep 2011 & $3.61 \pm 0.54$ & $0.16 \pm 0.02$ & $0.00 \pm 0.00$ & $0.00 \pm 0.00$ \\
Q3 & Sep 2011 & $0.96 \pm 0.14$ & $0.19 \pm 0.03$ & $0.29 \pm 0.04$ & $-2.28 \pm 0.48$ \\
Q2 & Sep 2011 & $1.07 \pm 0.16$ & $0.22 \pm 0.03$ & $0.57 \pm 0.09$ & $-3.07 \pm 0.65$ \\
\hline
Q0 & Jun 2014 & $1.04 \pm 0.16$ & $0.07 \pm 0.01$ & $0.00 \pm 0.00$ & $0.00 \pm 0.00$ \\
Q4 & Jun 2014 & $0.65 \pm 0.10$ & $0.11 \pm 0.02$ & $0.14 \pm 0.02$ & $-2.51 \pm 0.53$ \\
Q3 & Jun 2014 & $1.10 \pm 0.17$ & $0.16 \pm 0.02$ & $0.36 \pm 0.05$ & $-1.58 \pm 0.33$ \\
Q2 & Jun 2014 & $1.20 \pm 0.18$ & $0.25 \pm 0.04$ & $0.58 \pm 0.09$ & $5.04 \pm 1.06$ \\
\hline
Q0 & Jun 2015 & $1.03 \pm 0.16$ & $0.08 \pm 0.01$ & $0.00 \pm 0.00$ & $0.00 \pm 0.00$ \\
Q5 & Jun 2015 & $0.08 \pm 0.01$ & $0.09 \pm 0.01$ & $0.11 \pm 0.02$ & $-7.04 \pm 1.48$ \\
Q4 & Jun 2015 & $1.90 \pm 0.28$ & $0.19 \pm 0.03$ & $0.36 \pm 0.05$ & $-1.74 \pm 0.37$ \\
Q2 & Jun 2015 & $0.83 \pm 0.12$ & $0.30 \pm 0.04$ & $0.63 \pm 0.09$ & $5.45 \pm 1.15$ \\
\hline
Q0 & May 2017 & $0.32 \pm 0.05$ & $0.08 \pm 0.01$ & $0.00 \pm 0.00$ & $0.00 \pm 0.00$ \\
Q6 & May 2017 & $1.37 \pm 0.20$ & $0.34 \pm 0.05$ & $0.52 \pm 0.08$ & $2.51 \pm 0.53$ \\
Q5 & May 2017 & $0.53 \pm 0.08$ & $0.06 \pm 0.01$ & $0.59 \pm 0.09$ & $-10.82 \pm 2.24$ \\
Q4 & May 2017 & $1.34 \pm 0.20$ & $0.23 \pm 0.03$ & $0.72 \pm 0.11$ & $1.44 \pm 0.31$ \\
\hline
Q0 & Feb 2018 & $1.07 \pm 0.16$ & $0.05 \pm 0.01$ & $0.00 \pm 0.00$ & $0.00 \pm 0.00$ \\
Q6 & Feb 2018 & $0.69 \pm 0.10$ & $0.32 \pm 0.05$ & $0.54 \pm 0.08$ & $-1.67 \pm 0.35$ \\
Q5 & Feb 2018 & $0.84 \pm 0.13$ & $0.14 \pm 0.02$ & $0.80 \pm 0.12$ & $-3.78 \pm 0.80$ \\
Q4 & Feb 2018 & $1.17 \pm 0.18$ & $0.30 \pm 0.04$ & $0.84 \pm 0.12$ & $6.30 \pm 1.33$ \\
\hline
Q0 & Mar 2019 & $1.24 \pm 0.19$ & $0.04 \pm 0.01$ & $0.00 \pm 0.00$ & $0.00 \pm 0.00$ \\
Q6 & Mar 2019 & $0.22 \pm 0.03$ & $0.32 \pm 0.05$ & $0.51 \pm 0.08$ & $6.37 \pm 1.34$ \\
Q5 & Mar 2019 & $0.82 \pm 0.12$ & $0.15 \pm 0.02$ & $0.95 \pm 0.14$ & $-1.69 \pm 0.36$ \\
Q4 & Mar 2019 & $1.00 \pm 0.15$ & $0.41 \pm 0.06$ & $0.96 \pm 0.14$ & $6.07 \pm 1.28$ \\
\hline
\end{tabular}
\tablefoot{
(1): Component identifier.
(2): Observation epoch.
(3): Radio flux density.
(4): FWHM of the circular Gaussian.
(5): Projected distance from the core to the component.
\new{(6): Position angle (PA) of the circular Gaussian.}
Uncertainties are estimated as described in the text.
}
\label{tab:components}
\end{table*}

\subsection{Light-curve information}
\begin{figure}
    \centering
    \includegraphics[width=0.49\textwidth]{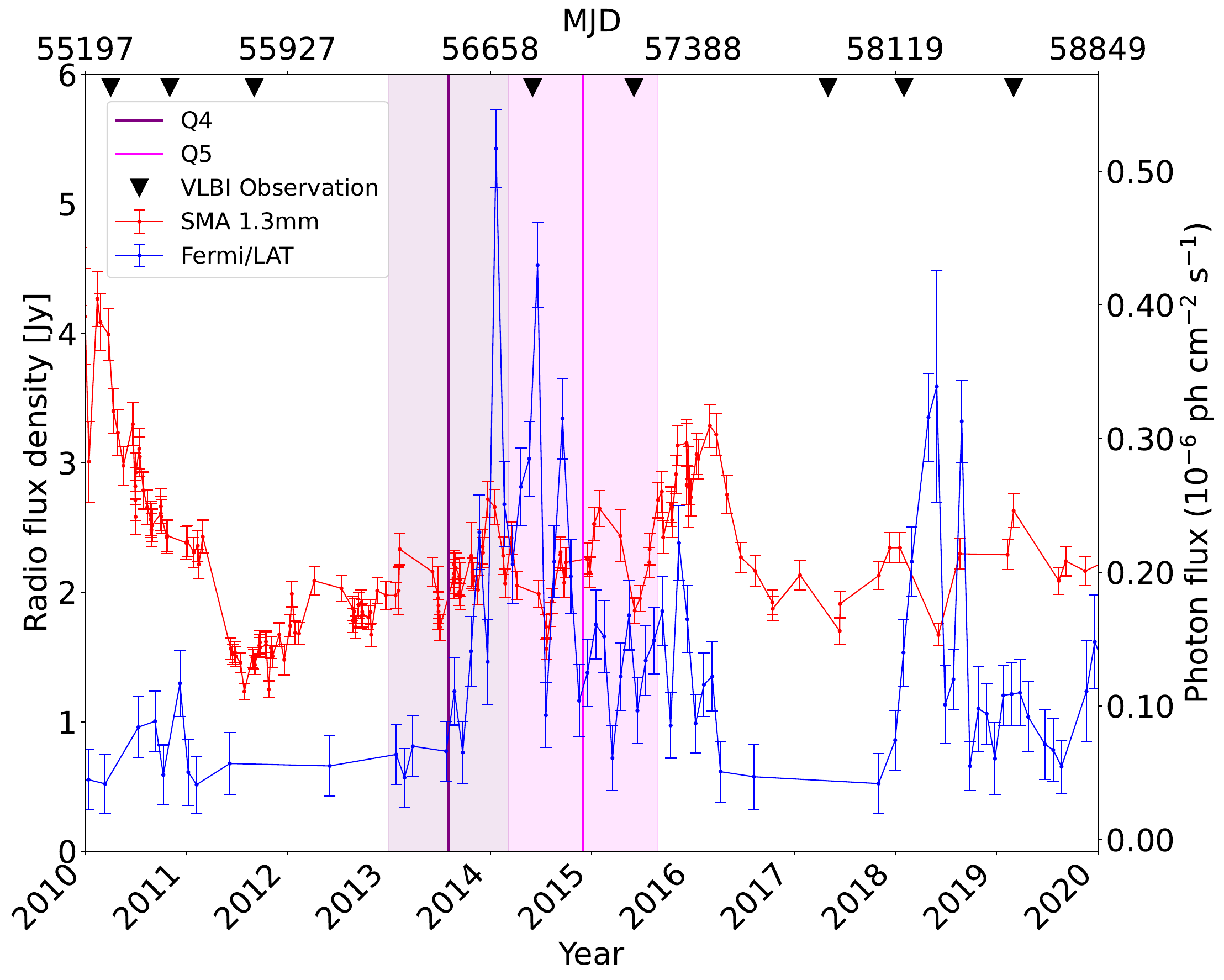}
    \caption{Radio and $\gamma$-ray light-curves of \N\, between 2010 and 2019. The red curve denotes data by SMA at 1.3\,mm, the blue denotes Fermi/LAT data, in which we adopt a monthly binning. The purple and magenta vertical lines indicate the jet component ejection epochs, and the shaded areas depict the uncertainty of the ejection epoch of component Q4 and Q5, respectively. The black triangles mark the dates of the VLBI observations used in this study. 
    }
    \label{fig:lightcurve}
\end{figure}

In order to investigate whether or not the emission of new jet components can be associated with the emission of high-energy photons, we utilised $\gamma$-ray and radio light-curve information from publicly available data by SMA and Fermi/LAT between 2010 and 2019. 
In Fig.~\ref{fig:lightcurve} we present the radio and $\gamma$-ray light-curves of \N, in which we adopted a monthly binning for the $\gamma$-rays. 
\new{Comparison with weekly and three-day binned light-curves confirms that the relevant flaring events are preserved and not introduced or suppressed by the chosen binning.}
% \new{in order to increase the signal-to-noise ratio and ensure a robust detection of the source over the full time range. 
% \new{We also examined the $\gamma$-ray light-curves using weekly and 3-day bins, however, they reveal the same flaring events and do not show additional flares that are absent in the monthly binned data. 
% Therefore, the relevant flaring activities discussed in this work are preserved using the monthly binning and are not sensitive to the adopted binning. }
From 2010 to mid 2012 the radio flux density shows a flare in the beginning of 2010, with a maximum of $(4.27\pm0.21)~\mathrm{Jy}$ \new{(February 2010)}, after which the flux density decreases until 2015. 
We defined a flare as the radio flux density exceeding two times the quiescent level ($\sim2~\mathrm{Jy}$). 
As for the $\gamma$-rays, we defined a $\gamma$-ray flare as the photon flux, which exceeds five times the quiescent level of $\sim0.05\times10^{-6}~\mathrm{ph~ cm^{-2}~s^{-1}}$. 
We observe a $\gamma$-ray flare starting in January 2014 (with peaks in June and September) with a maximum of $(0.52\pm0.03)\times10^{-6}~\mathrm{ph~cm^{-2}~s^{-1}}$.  
In May 2018, we observe a second, less prominent $\gamma$-ray flare, occurring a few months after the VLBI observations in that year with a maximum of $(0.34\pm0.03)\times10^{-6}~\mathrm{ph~ cm^{-2}~s^{-1}}$. 
The interpretation and implications of these observations, together with the jet kinematics, are discussed in Sect.~\ref{Sec:Discussion}. 

\subsection{Polarimetric analysis and EVPA geometry}
\begin{figure*}
    \centering
    \includegraphics[width=\textwidth]{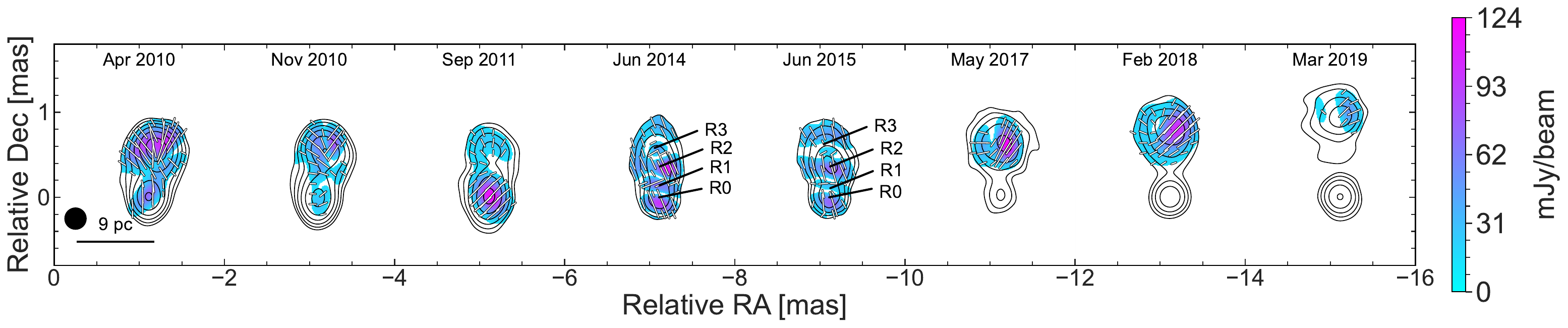}
    \caption{Linear polarisation images of \N\, of all epochs at 43\,GHz. 
    The total intensity flux density is represented by the black contours, using the contour levels of 2, 4, 8, 16, 32, and 64\% of the total intensity peak of each map, the linear polarisation is represented by the colour scale using a cut-off at $ 6\sigma_\textrm{P}$ for (with $\sigma_\textrm{P} = 2.3\,\textrm{mJy/beam}$), and the white ticks represent the EVPAs. 
    The marked regions R0-R3 denote the regions of enhanced polarised emission. 
    The black circle in the bottom left corner denotes the convolving, circular beam size of 0.25\,mas for all epochs and the black dash in the bottom left corner denotes the projected distance of $9\,\mathrm{pc}$ corresponding to $20000\,R_\mathrm{S}$. }
    \label{fig:Polarisationmaps}
\end{figure*}

Furthermore, we employed polarimetric information of \N\, from which we can directly probe the magnetic field configuration from the EVPAs. 
Figure~\ref{fig:Polarisationmaps} depicts the linear polarisation information of \N\, at 43\,GHz across all epochs.  
We find that the polarisation signal in the first three epochs (2010-2011) is high near the VLBI core and jet region. 
After 2014 and 2015, the prominent polarisation signal is detected only in the extended jet region. 
Across the time span from 2010 until 2019, we find that the most prominent linear polarisation detection shifts from both the core and jet region to being solely in the jet, with a prominent transition in 2014 and 2015. 
In these epochs, four regions show enhanced linear polarisation, which are the core (R0) and regions located north of the core (R1, R2, R3), while moderate polarisation is observed further downstream in the jet. 
All regions of enhanced emission (R0-R3) coincide with the positions of the \textsc{model-fit} components. 
In the last three epochs, however, we only detect linear polarisation in the extended jet region. 

The integrated linear polarisation shows increased emission during the epochs of enhanced radio and $\gamma$-ray emission (April 2010, June 2014 and 2015, and February 2018), reaching values of $0.30-0.36~\mathrm{Jy}$, compared to lower linear polarisation of $\sim0.14~\mathrm{Jy}$ observed during quiescence. 
\new{Similarly, the integrated fractional polarisation signal has increased during the $\gamma$-ray flares in 2014 and 2018, reaching values of up to $\sim10\%$.} 
The EVPAs in Fig.~\ref{fig:Polarisationmaps} show a relative change of orientation with respect to the jet around the VLBI core, i.e., the direction of the EVPAs is parallel to the propagation of the moving components. 
% \ld{\textit{Add information about the average integrated polarised flux from core region and extended jet. 
% Average over all EVPA in these areas. }}
\new{In the downstream jet the EVPAs are oriented like a fan, that is, their orientation is parallel to the jet flow and appears to spread outward with respect to the jet axis. 
As a consequence, the magnetic field is expected to be directed perpendicular to the propagation of the jet, which provides evidence for a toroidal magnetic field configuration in the downstream jet. 
}
% However, in the core region we can not make strong assumptions on the magnetic field direction, as the opacity effects may affect the observed EVPA and the perpendicular relation between the EVPA and the magnetic field direction does not necessarily hold. }

%--------------------------------------------------------------------
%--------------------------------------------------------------------
%--------------------------------------------------------------------
%--------------------------------------------------------------------
%--------------------------------------------------------------------
%--------------------------------------------------------------------
\section{Discussion\label{Sec:Discussion}}
The multi-epoch high angular resolution analysis of \N\, presented in this work reveals new insights into the kinematic behaviour of its jet, and its connection to the observed $\gamma$-ray activity. 
Combining VLBI total intensity and linear polarisation imaging with radio and $\gamma$-ray light-curves, we investigate the temporal relation between the ejection of new jet components and high-energy emission. 
% -----------------------------
%       Kinematic Analysis
% _____________________________
\subsection{Kinematic analysis\label{subsec:Kin}}
We traced distinct emission features across multiple epochs using Stokes~I images and determined their projected velocities under the assumption of linear motion. 
We note that, while previous studies applied non-linear models to fit the component kinematics~\citep[e.g., ][]{Agudo2007, Molina2014}, their estimated projected velocities are in agreement with the velocities presented in this work. 
Thus, even though our assumed linear motion is a first-order approximation, adopted to minimise the number of free parameters, given the limited available VLBI epochs and positional uncertainties, this approach allows for a robust, quantitative comparison with previous studies. 
Note here, that projected velocities depend on the identification of the components and can differ between studies. 
While the VLBI core is often associated with the brightest feature in the Stokes~I map at one edge of the jet, this is not universally applicable to all sources and epochs. 
In this work, we identified the core of \N\, as the southernmost component, which does not always coincide with the brightest feature in the Stokes~I map (see e.g. April 2010 in Fig.~\ref{fig:Modelfitmaps}). 
This identification is motivated by the component’s persistent location at the south of the jet, its compact morphology, and its role as the apparent origin of downstream emission features extending north across all epochs. 
For consistency within our kinematic analysis, we adapt a fixed geometric reference point. 

\new{
We note that the kinematic interpretation presented here depends on the adopted cross-identification of the \textsc{model-fit} components across all epochs. 
Given the complex morphology of the inner jet and the limited VLBI observations, the identification of individual components is not unique. 
In an alternate scenario components Q3, Q4 (June 2015), and Q6 (May 2017) could also correspond to the same physical feature. 
Similarly, components Q2 and Q4 (May 2017) could be interpreted as the same component. 
These alternative identifications would modify the inferred kinematics and interpretation. 
However, the components detected in the later epochs (2017-2019) must have been ejected at some earlier time and should therefore be connected to preceding jet features. 
% However, independent of the adopted cross-identification, the components detected in the later epochs (2017-2019) must have been ejected at some earlier time and would therefore be connected to preceding jet features. 
% In this context, the new components identified in June 2014 and June 2015 provide plausible candidates for such earlier ejecta, as they also appear during the period of enhanced $\gamma$-ray activity and can be naturally connected to the downstream components observed in subsequent epochs. 
The new components identified in June 2014 and June 2015 provide plausible candidates for such earlier ejecta, as they also appear during the period of enhanced $\gamma$-ray activity and can be naturally connected to the downstream components observed in subsequent epochs. 
% Thus, although alternative cross-identifications cannot be excluded, the adopted identification provides a coherent evolutionary picture as it provides the most consistent description of the component in trajectories, flux density, and FWHM across all epochs. 
}

% Although alternative cross-identifications cannot be excluded, 
\new{Overall, our adopted identification is characterised by a coherent evolution in trajectories, flux density, and FWHM across all epochs, establishing its robustness.
Furthermore, the projected velocities derived from this cross-identification are consistent with previous kinematic studies of \N, such as the proper motions reported by \cite{Agudo2007} and \cite{Molina2014}, supporting the adopted cross-identification.
}
% -----------------------------
%       Connection gamma <-> ejection epoch
% _____________________________
\subsection{Radio and $\gamma$-ray emission}
% \subsection{$\gamma$-ray emission}
% While the radio flux density is high and decreasing
\new{We further considered light-curve information in order to investigate a possible connection between enhanced radio and $\gamma$-ray emission and the ejection of new jet components. }
\new{During the first three epochs, from April 2010 until September 2011, the radio flux density shows a gradual decline, while significant total flux density and linear polarisation is detected in both the core region and downstream of the core (Fig.~\ref{fig:Modelfitmaps} and \ref{fig:Polarisationmaps}). 
Enhanced total intensity and linear polarisation on short time scales traces regions of ordered magnetic fields and relativistically boosted synchrotron emission~\citep[see][for a review]{Raiteri2024arx}, suggesting that a substantial fraction of the observed radio emission originates not only from the core, but also in the downstream jet. 
After the radio flare, the increased total flux density is shifting from the extended jet closer to the core region. 
}
% \ld{During the radio flare in February 2010 the total intensity is high in the core and in the extended jet. 
% Under the assumption that ordered magnetic fields trace regions of enhanced radio emission, we can conclude that the radio emission may arise from the core and the extended jet. 
% This can be indicative for \textit{the radio emission is originating from the core and the downstream jet~\citep{}. } 

Two new emission features (Q4 and Q5) emerge close to the core region during the enhanced $\gamma$-ray activity in 2014, whereas during the $\gamma$-ray quiescence previously ejected components (Q1, Q2, Q3) exhibit relatively slow apparent motion. 
To investigate the temporal relationship between component ejection and high-energy emission and which event happens first (i.e., emission of new components following $\gamma$-ray flares or vice versa), we utilised the extrapolation of the linear fits. 
The extrapolated ejection epochs of components Q4 and Q5 are marked with vertical lines and the uncertainty of the fits are shown by the shaded areas (see Fig.~\ref{fig:lightcurve}). 
Our results indicate that component Q4 is ejected \new{five months} before the onset $\gamma$-ray flares observed in 2014, while component Q5 appears to be ejected $\sim1~\mathrm{yr}$ after the enhanced emission. 
However, given the uncertainties of the ejection epochs, we can not robustly conclude whether the ejection of component Q4 is ahead of the $\gamma$-ray flare, or if it appears contemporaneously.
The overlap of the ejection of Q4 and Q5, however, is happening close after the first $\gamma$-ray flare in 2014. 
\cite{Jorstad2001} studied the temporal coincidence of $\gamma$-ray flares and the ejection of superluminal components from EGRET blazars and found, that $\gamma$-ray flares are followed by the ejection of new superluminal VLBI components.  
A similar behaviour is observed for component Q5 presented in this work, indicating that its ejection may be associated with the preceding $\gamma$-ray flare. 
\new{We note here, that the estimated time of ejection refers to the time when this component is first detectable at 43~GHz and does not refer to the actual emission from the core. 
% However, previous studies showed that the core-shift at higher frequencies is not significant compared to the uncertainties of our linear fit~\citep{Pushkarev2012}, thus the time of detection is expected to be close the time of ejection.
% However, previous studies showed that the core-shift at higher frequencies is not significant compared to the uncertainties of our linear fit~\citep{Pushkarev2012}, thus the time of detection of the component is expected to be close to the time of ejection.
However, previous studies showed that the core-shift at higher frequencies is not significant compared to the uncertainties of our linear fit~\citep{Pushkarev2012}. Therefore, the distance between the observed radio core and the black hole is expected to be negligible, implying that the time of detection of the component is close to its time of ejection. 
}
% -----------------------------
%       origin gamma emission
% _____________________________
The timing of the $\gamma$-ray emission in \N\, suggests that the high-energy emission may not originate close to the SMBH but rather at some distance downstream in the jet, where the newly ejected features interact with the surrounding jet flow or undergo further particle acceleration~\citep{Marscher2008, Debbrecht2026}. 
Such a scenario is consistent with studies that place the $\gamma$-ray emission site downstream of the VLBI core in AGN jets~\citep[such as seen in 3C\,84 and other blazars;][]{Hodgson2021, GP2026, Tavares2011}. 
% \ld{\textit{Add information about fractional polarisation in core region and in extended jet area; compare value in these areas and make quantitative analysis of emission region of high-energy emission. }}
The $\gamma$-ray flares might be produced by inverse Compton (IC) scattering downstream the jet rather than in the immediate vicinity to the central black hole~\citep{Jorstad2001, Kovalev2009, Jo2026}. 
Moreover, the $\gamma$-ray emission results primarily from IC scattering, in which seed photons originate from the broad-line region~\citep[][]{Ramakrishnan2015, Zhou2018}. 
Another AGN pointing with a similar viewing angle towards our LOS is the BL-Lac object OJ~287~\citep[$\theta\approx3^\circ-8^\circ$; ][]{Traianou2025, OJ287EHT2026}, in which the high-energy emission appears to originate from either the VLBI core, from a feature downstream the jet, or from both sites~\citep{Hodgson2017}. 
In order to locate the high-energy emission site in more detail, we utilise linear polarisation information. 
\subsection{Linear polarisation}
% -----------------------------
%       Radio emission and Polarisation
% _____________________________
% During the first three epochs, from April 2010 until September 2011, the radio flux density shows a gradual decline, while significant linear polarisation is detected in both the core region and downstream of the core (Fig.~\ref{fig:Polarisationmaps}). 
% Enhanced linear polarisation on short time scales traces regions of ordered magnetic fields and relativistically boosted synchrotron emission~\citep[see][for a review]{Raiteri2024arx}, suggesting that a substantial fraction of the observed radio emission originates not only from the core, but also in the downstream jet. 

% -----------------------------
%       lin. polarisation information
% _____________________________
During periods of enhanced radio and $\gamma$-ray activity (see Fig.~\ref{fig:lightcurve}), the integrated linear polarisation is increased, further supporting the interpretation that the enhanced emission is associated with regions of ordered magnetic fields and relativistically boosted synchrotron radiation in the jet. 
The locations of enhanced linear polarisation is locally coinciding with the positions of the \textsc{model-fit} components and contemporaneous with the $\gamma$-ray flare, indicating a connection between the high-energy emission and the increased polarised emission within \N. 
Similarly to the case of the blazar OJ~287, where a $\gamma$-ray flare with comparable intensity to the peak in \N\, \citep[$\sim0.5\times10^{-6}~\mathrm{ph~ cm^{-2}~s^{-1}}$;][]{Agudo2011} appears to originate at the site of high linear polarisation, we observed increased linear polarisation. 
This support the hypothesis, that the $\gamma$-ray flares in \N\, might be produced on the sites of intense linear polarised emission.
Following the $\gamma$-ray flares in 2014, linear polarisation is detected predominantly in the downstream jet for the epochs from 2017 to 2019, while the core region appears weakly polarised or unpolarised. 
However, the increased linear polarisation emission in 2018 during the second, less prominent $\gamma$-ray flare may be related to the emission site of the $\gamma$-rays, as previously proposed by \cite{Agudo2011} and \cite{GP2026}, and might be produced by IC scattering, either produced by synchrotron self-Compton (SSC) mechanism or IC scattering of infrared radiation from a hot, dusty torus at the position of enhanced polarised emission. 
We also consider a ring of fire scenario, in which a relativistic plasma blob propagates along the spine of the blazar passing through a ring that is emitting synchrotron emission~\citep{MacDonald2015, Traianou2026arXiv}. 
For this source, however, there is no optical light-curve information available to test this model. 
Under the assumption that enhanced linear polarisation traces regions associated with 
$\gamma$‑ray flaring activity, the observed polarisation behaviour supports a scenario in which the high‑energy emission originates downstream of the central engine and is linked to evolving jet features.

% -----------------------------
%       EVPAs/ Magnetic field config. 
% _____________________________
Finally, from the morphology of the fan-like orientation of EVPAs in \new{the downstream jet} we found evidence for a toroidal magnetic field in the jet of \N. 
This interpretation is consistent with the results of \cite{Molina2014}, who reported a toroidal magnetic field configuration utilising polarimetric information and \cite{Livingston2025}, who inferred a combined helical and toroidal magnetic field in the innermost jet based on their estimated rotation measure~\citep{Broderick2010, Gabuzda2015}. 
Such magnetic field configurations are naturally expected in magnetically launched relativistic jets and may play an important role in governing both jet dynamics and high-energy emission~\citep{Kramer2021}.
\new{While the observed EVPA structure can be indicative of a toroidal magnetic field, it is also possible that this morphology is produced by shocks~\citep{Cawthorne2006, Poetzl2021}. 
In particular, simulations of synchrotron emission from oblique shocks in relativistic jets show that compression of the magnetic field at the shock front can produce characteristic polarisation patterns, including spine–sheath EVPA structures depending on the viewing geometry and upstream magnetic field configuration~\citep[e.g.][]{Cawthorne2006}. 
Therefore, while the observed EVPA structure in \N\, is consistent with the presence of a toroidal magnetic field, shock-induced compression of the magnetic field cannot be excluded as an alternative explanation.
Furthermore, in optically thick regions the observed EVPAs are not necessarily perpendicular to the magnetic field direction. 
Hence, we cannot draw a strong conclusion about the magnetic field orientation in the core region, as opacity effects might influence the observed EVPAs. 
}

\section{Conclusions \label{Sec:Conclusion}}
We conducted a multi‑epoch VLBI analysis of the innermost jet of \N\, at 43\,GHz, examining the kinematics of individual emission features and their relation to $\gamma$‑ray activity and linear polarisation. The main results of this study are summarised as follows:
\begin{itemize}
    \item We traced multiple emission features in the innermost jet of \N\, using multi-epoch VLBI observations and derived their projected velocities under the assumption of linear motion, finding projected velocities between $0.036\pm0.015~\mathrm{mas/yr}$ and $0.243\pm0.025~\mathrm{mas/yr}$ (corresponding to $1.06 \pm 0.44\,c$ and $7.13 \pm 0.73\,c$). 
    \item We find a connection between enhanced $\gamma$-ray activity and the emergence of new emission features Q4 and Q5, showing a temporal association with a preceding $\gamma$-ray flare.
    \item The event of the $\gamma$-ray flares in 2014 and the extrapolated component ejection epochs of Q4 and Q5 supports a scenario in which the high-energy emission may be produced at locations downstream from the black hole.
    \item Linear polarisation is detected in both the core and downstream jet, with enhanced polarised emission during periods of increased $\gamma$-ray activity reaching values of $0.30-0.36~\mathrm{Jy}$; following the major flares, the polarised emission shifts predominantly to the downstream jet, supporting a scenario in which the high-energy emission site is associated with evolving jet features rather than a stationary core. 
    \item The observed morphology of the polarised emission provides evidence for a toroidal magnetic field configuration in the inner jet region of \N.
\end{itemize}
Our results support a scenario in which $\gamma$-ray flaring activity in \N\, is closely linked to the dynamical evolution of the relativistic jet and its magnetic field structure.
This study highlights the importance of combining multi‑epoch VLBI kinematics and polarimetric information with high‑energy observations to constrain the location and physical origin of 
$\gamma$‑ray emission in blazars.

\begin{acknowledgements}
\new{We would like to thank the anonymous referee for their constructive comments, which improved our work.} 
We thank I. Liodakis for valuable comments and insightful discussions. 
This publication acknowledges project M2FINDERS, which is funded by the European Research Council (ERC) under the European Union’s Horizon 2020 research and innovation programme (grant agreement no. 101018682). 
I.~A. acknowledges financial support from the grant CEX2021-001131-S funded by MCIN/AEI/10.13039/501100011033 to the Instituto de Astrof\'isica de Andaluc\'ia-CSIC and from  MICIN grant PID2022-139117NB-C44. 
E.~R. and J.~A.~Zensus are supported by the Deutsche Forschungsgemeinschaft (DFG, German Research Foundation) as part of the DFG Research Unit FOR5195, project number 443220636. 
The European VLBI Network is a joint facility of independent European, African, Asian, and North American radio astronomy institutes. 
Scientific results from data presented in this publication are derived from the following EVN project code: GA032.
This study makes use of VLBA data from the VLBA-BU Blazar Monitoring Program (BEAM-ME and VLBA-BU-BLAZAR; \href{http://www.bu.edu/blazars/BEAM-ME.html}{http://www.bu.edu/blazars/BEAM-ME.html}), funded by NASA through the Fermi Guest Investigator Program.  
The Submillimeter Array is a joint project between the Smithsonian Astrophysical Observatory and the Academia Sinica Institute of Astronomy and Astrophysics and is funded by the Smithsonian Institution and the Academia Sinica. 
We recognize that Maunakea is a culturally important site for the indigenous Hawaiian people; we are privileged to study the cosmos from its summit. 
This research is based in part on observations obtained with the 100-m telescope of the MPIfR at Effelsberg, observations carried out at the IRAM 30-m telescope operated by IRAM, which is supported by INSU/CNRS (France), MPG (Germany) and IGN (Spain), observations obtained with the Yebes 40-m radio telescope at the Yebes Observatory, which is operated by the Spanish Geographic Institute (IGN, Ministerio de Transportes, Movilidad y Agenda Urbana), and observations supported by the Green Bank Observatory, which is a main facility funded by the NSF operated by the Associated Universities. We acknowledge support from the Onsala Space Observatory national infrastructure for providing facilities and observational support. The Onsala Space Observatory receives funding from the Swedish Research Council through grant no. 2017-00648. This publication makes use of data obtained at the Metsähovi Radio Observatory, operated by Aalto University. 

\end{acknowledgements}

\bibliographystyle{aa} % style aa.bst
\bibliography{aanda} 
\end{document}